\documentclass[12pt]{iopart}

\usepackage{graphicx}
\usepackage{amssymb}

\begin{document}
\title[Incoherent superconductivity well above $T_{\rm c}$ in high-$T_{\rm c}$ cuprates]{Incoherent superconductivity well above $T_{\rm c}$ in high-$T_{\rm c}$ cuprates -- harmonising the spectroscopic and thermodynamic data}
\author{J G Storey$^{1,2}$}

\address{$^1$ Robinson Research Institute, Victoria University of Wellington, P.O. Box 600, Wellington, New Zealand}
\address{$^2$ MacDiarmid Institute, Victoria University of Wellington, P.O. Box 600, Wellington, New Zealand}
\ead{james.storey@vuw.ac.nz}

\begin{abstract}
Cuprate superconductors have long been known to exhibit an energy gap that persists high above the superconducting transition temperature ($T_{\rm c}$). Debate has continued now for decades as to whether it is a precursor superconducting gap or a pseudogap arising from some competing correlation. Failure to resolve this has arguably delayed explaining the origins of superconductivity in these highly complex materials. Here we effectively settle the question by calculating a variety of thermodynamic and spectroscopic properties, exploring the effect of a temperature-dependent pair-breaking term in the self-energy in the presence of pairing interactions that persist well above $T_{\rm c}$. We start by fitting the detailed temperature-dependence of the electronic specific heat and immediately can explain its hitherto puzzling field dependence.
Taking this same combination of pairing temperature and pair-breaking scattering we are then able to simultaneously describe in detail the unusual temperature and field dependence of the superfluid density, tunneling, Raman and optical spectra, which otherwise defy explanation in terms a superconducting gap that closes conventionally at $T_{\rm c}$. These findings demonstrate that the gap above $T_{\rm c}$ in the overdoped regime likely originates from incoherent superconducting correlations, and is distinct from the competing-order ``pseudogap'' that appears at lower doping.
\end{abstract}
\pacs{74.72.Gh, 74.25.Bt, 74.25.nd, 74.55.+v}
\vspace{2pc}
\noindent{\it Keywords}: cuprates, scattering, specific heat, superfluid density, Raman spectroscopy, tunneling, optical conductivity

\submitto{\NJP}
\maketitle

\section{Introduction}

\begin{figure}
\centering
\includegraphics[width=13cm]{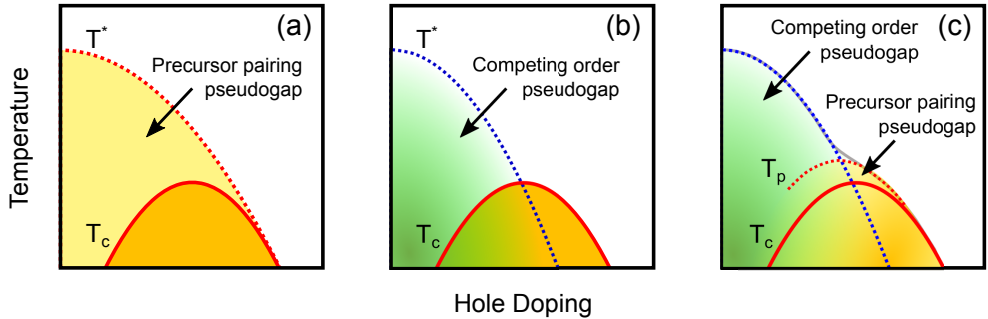}
\caption{
Candidate phase diagrams for the hole-doped cuprates. For simplicity only the superconducting and pseudogap phases are shown. (a) Precursor pairing scenario. (b) Competing or coexisting order parameter scenario. (c) Combined phase diagram proposed in this work.
} 
\label{TSTARFIG}
\end{figure}

A prominent and highly debated feature of the high-$T_{\rm c}$ cuprates is the presence of an energy gap at or near the Fermi level which opens above the observed superconducting transition temperature. It is generally known as the ``pseudogap''. Achieving a complete understanding of the pseudogap is a critical step towards the ultimate goal of uncovering the origin of high-temperature superconductivity in these materials. For example, knowing where the onset of superconductivity occurs sets limits on the strength of the pairing interaction. The community has long been divided between two distinct viewpoints\cite{PG3}. These can be distinguished by the doping dependence of the so-called $T^*$ line\cite{OURWORK1}, the temperature below which signs of a gap appear. The first viewpoint holds that the pseudogap represents precursor phase-incoherent superconductivity or ``pre-pairing''. In this case of a single $d$-wave gap the pseudogap opens at $T^*$ and evolves into the superconducting gap below $T_{\rm c}$. The underlying Fermi surface is a nodal-metal, appearing as an arc due to broadening processes\cite{FERMIARCS2,REBER2012}. Here the $T^*$ line merges smoothly with the $T_{\rm c}$ dome on the overdoped side (see figure~\ref{TSTARFIG}(a)). The second viewpoint is that the pseudogap arises from some as yet unidentified competing and/or coexisting order. In this two-gap scenario the pseudogap is distinct from the superconducting gap with a different momentum dependence, likely resulting from Fermi surface reconstruction\cite{STOREYHALL}. The $T^*$ line in this case bisects the $T_{\rm c}$ dome and need not be a transition temperature in the thermodynamic sense or ``phase transition'', where it would instead mark a crossover region defined by the energy of a second order parameter given by $E_g \approx 2k_{\rm B}T^*$ (see figure~\ref{TSTARFIG}(b)). Ironically, the multitude of different techniques employed to study the pseudogap has lead to much confusion over the exact form of the $T^*$ line.

However, an alternative picture is beginning to emerge that encompasses both viewpoints (see figure~\ref{TSTARFIG}(c)). Small superconducting coherence lengths in the high-$T_{\rm c}$ cuprates give rise to strong superconducting fluctuations that are clearly evident in many techniques.  Thermal expansivity\cite{MEINGAST}, specific heat\cite{WEN2009,TALLONFLUCS}, resistivity\cite{ALLOUL2010}, Nernst effect\cite{XU,WANG2006,CHANG}, THz conductivity\cite{BILBRO2011}, IR conductivity\cite{DUBROKA} and Josephson effect\cite{BERGEAL} measurements show that although the fluctuation regime persists as high as 150 K\cite{TALLONFLUCS}, it is confined to a narrower region above $T_{\rm c}$ and does not track the $T^*$ line\cite{ALLOUL2010,WANG2006,DUBROKA,TALLONFLUCS} which extends to much higher temperatures at low doping. An effective superconducting gap feature associated with these fluctuations which tails off above $T_{\rm c}$ can be extracted from the specific heat \cite{TALLON2013}. And pairing gaps above $T_{\rm c}$ have been detected by scanning tunneling microscopy in this temperature range\cite{GOMES}. 
Evidence for a second energy scale, which from here will be referred to specifically as the pseudogap, includes a downturn in the normal-state spin susceptibility\cite{WILLIAMS97,ENTROPYDATA2} and specific heat\cite{ENTROPYDATA2}, a departure from linear resistivity\cite{ANDO,TSTAR,DAOU2009}, and a large gapping of the Fermi surface at the antinodes by angle-resolved photoemission spectroscopy (ARPES)\cite{TANAKA,LEE,VISHIK2010,MATT}. The opening of the pseudogap at a critical doping within the $T_{\rm c}$ dome can be inferred from an abrupt drop in the doping dependence of several properties. These include the specific heat jump at $T_{\rm c}$\cite{ENTROPYDATA2}, condensation energy\cite{ENTROPYDATA2}, zero-temperature superfluid density\cite{ENTROPYDATA2,BERNHARD}, the critical zinc concentration required for suppressing superconductivity\cite{TALLON4}, zero-temperature self-field critical current\cite{NAAMNEH}, and the Hall number\cite{BADOUX}, most of which represent ground-state  properties. The last signals a drop in carrier density from $1+p$ to $p$ holes per Cu, and can be explained in terms of a reconstruction from a large to small Fermi surface\cite{STOREYHALL}. At or above optimal doping the pseudogap becomes similar or smaller in magnitude than the superconducting gap and, since many techniques return data that is dominated by the larger of the two gaps, it has been historically difficult to determine which gap is being observed. In this work it will be demonstrated explicitly that in this doping range it is in fact the superconducting gap persisting above $T_{\rm c}$ that is being observed, thereby ending the confusion over the shape of the $T^*$ line.

This work was inspired by two recent studies. The first by Reber \textit{et al}.\cite{REBER} fitted the ARPES-derived tomographic density of states using the Dynes equation\cite{DYNES} 
\begin{equation}
I_{TDoS}=\rm{Re}\frac{\omega-i\Gamma_{\rm s}}{\sqrt{(\omega-i\Gamma_{\rm s})^2-\Delta^2}}
\label{eq:Dynes}
\end{equation}
to extract the temperature dependence of the superconducting gap $\Delta$ and the pair-breaking scattering rate $\Gamma_{\rm s}$.
They found that $\Delta$ extrapolates to zero above $T_{\rm c}$ while $\Gamma_{\rm s}$ increases steeply near $T_{\rm c}$. They also found that $T_{\rm c}$ occurs when $\Delta = 3\Gamma_{\rm s}$. Importantly, these parameters describe the filling-in behaviour of the gapped spectra with temperature (originally found in tunneling experiments e.g. \cite{MIYAKAWA,MATSUDA,OZYUZER} and also inferred from specific heat and NMR\cite{WILLIAMSPRB98}), as opposed to the closing behaviour expected if $\Delta$ was to close at $T_{\rm c}$ in the presence of constant scattering.

The second study, by Kondo \textit{et al}.\cite{KONDO2014}, measured the temperature dependence of the spectral function around the Fermi surface using high-resolution laser ARPES. This was fitted using the phenomenological self-energy proposed by Norman \textit{et al}.\cite{NORMAN}
\begin{equation}
\Sigma(\textbf{k},\omega)=-i\Gamma_{\rm{single}}+\frac{\Delta^2}{\omega+\xi(\textbf{k})+i\Gamma_{\rm{pair}}}
\label{eq:ansatz}
\end{equation}
where $\xi(\textbf{k}$) is the energy-momentum dispersion, $\Gamma_{\rm{single}}$ is a single-particle scattering rate and $\Gamma_{\rm{pair}}$ is a pair-breaking scattering rate. The gap is well described by a $d$-wave BCS temperature dependence with an onset temperature $T_{\rm{pair}}$ above the observed $T_{\rm c}$. $\Gamma_{\rm{pair}}$ increases steeply near $T_{\rm c}$, with $T_{\rm c}$ coinciding with the temperature where $\Gamma_{\rm{pair}}=\Gamma_{\rm{single}}$. The aim of the present work is to investigate whether other experimental properties are consistent with this phenomenology. The approach is to fit the bulk specific heat using \ref{eq:ansatz} then, using the same parameters, calculate the superfluid density, tunneling and Raman spectra, and optical conductivity. To reiterate, the focus here is the overdoped regime near $T_{\rm c}$ where the pseudogap and subsidiary charge-density-wave order are absent\cite{BADOUXPRX}.

\section{Results}
\subsection{Specific Heat}
The Green's function with the above self-energy (\ref{eq:ansatz}) is given by
\begin{equation}
G(\xi,\omega)=\frac{1}{\omega-\xi+i\Gamma_{\rm{single}}-\frac{\Delta^2}{\omega+\xi+i\Gamma_{\rm{pair}}}}
\label{eq:G}
\end{equation}
The superconducting gap is given by $\Delta = \Delta_0\delta(T)\cos 2\theta$, where $\Delta_0=2.14k_{\rm B}T_{\rm p}$ and $\delta(T)$ is the $d$-wave BCS temperature dependence. $\theta$ represents the angle around the Fermi surface relative to the Brillouin zone boundary and ranges from 0 to $\pi/2$.
The density of states $g(\omega)$ is obtained by integrating the spectral function $A(\xi,\omega)=\pi^{-1}\rm{Im}G(\xi,\omega)$
\begin{equation}
g(\omega) = \int{A(\xi,\omega)d\xi d\theta}
\label{eq:DOS}
\end{equation}
The electronic specific heat coefficient $\gamma(T)=\partial S/\partial T$ is calculated from the entropy
\begin{equation}
S(T) = -2k_{\rm B}\int{[f\ln f + (1-f)\ln (1-f)]g(\omega)d\omega}
\label{eq:gamma}
\end{equation}
where $f$ is the Fermi distribution function. The temperature dependence of $\Gamma_{\rm{pair}}$ is extracted by using it as an adjustable parameter to fit specific heat data under the following assumptions: i) the superconducting gap opens at $T_{\rm p}$ = 120 K, at the onset of superconducting fluctuations; and ii) A linear-in-temperature $\Gamma_{\rm{single}}$ ranging from 5 meV at 65 K to 14 meV at 135 K, similar to values reported by Kondo \textit{et al}.\cite{KONDO2014}. A difficulty in applying this approach over the whole temperature range is that the $T$-dependence of the underlying normal-state specific heat $\gamma_{\rm n}$ must be known. Therefore attention will be focused close to $T_{\rm c}$ on Bi$_2$Sr$_2$CaCu$_2$O$_{8+\delta}$ data\cite{ENTROPYDATA2} with a doping of 0.182 holes/Cu, where $\gamma_{\rm n}$ can be taken to be reasonably constant. In practice the quantity fitted is the dimensionless ratio of superconducting- to normal-state entropies $S_{\rm s}(T)/S_{\rm n}(T)$.
\begin{figure}
\centering
\includegraphics[width=13cm]{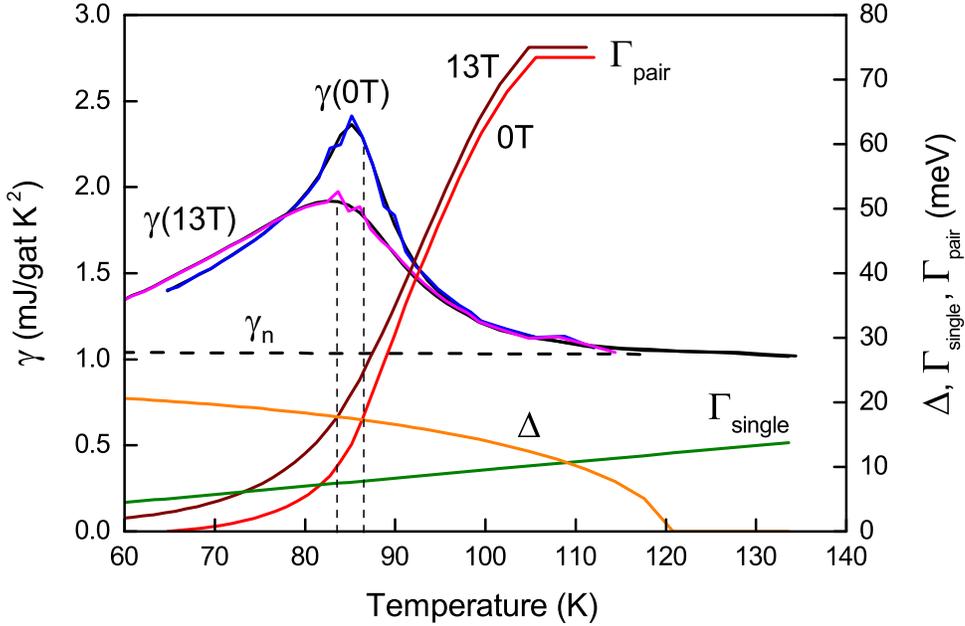}
\caption{
Fits (blue and magenta lines) made to the electronic specific heat of slightly overdoped Bi$_2$Sr$_2$CaCu$_2$O$_{8+\delta}$ ($p$ = 0.182 holes/Cu) at zero and 13 T fields (black lines) using the self-energy given by \ref{eq:ansatz}. $\Gamma_{\rm{pair}}$ is the adjustable parameter and $\Delta(T)$ and $\Gamma_{\rm{single}}(T)$ are assumed.
} 
\label{FITFIG}
\end{figure}

Fits and parameters are shown in figure~\ref{FITFIG} for data measured at zero and 13 T applied magnetic field. $\Gamma_{\rm{pair}}$ increases steeply near $T_{\rm c}$ in a very similar manner to the scattering rates found from the ARPES studies mentioned above. No particular relationship between $\Gamma_{\rm{pair}}$, $\Gamma_{\rm{single}}$ and $T_{\rm c}$ is observed, however the peak of the specific heat jump occurs when $\Gamma_{\rm{pair}} = \Delta$. In other words, once the pair-breaking becomes of the order the superconducting gap the entropy changes less rapidly with temperature, which intuitively makes sense. This appears to differ significantly with the result $\Delta(T_{\rm c})=3\Gamma_{\rm s}(T_{\rm c})$ from Reber \textit{et al}.\cite{REBER}, but note that fitting with the Dynes equation (\ref{eq:Dynes}) returns a smaller scattering rate $\Gamma_{\rm s}$ equal to the average of $\Gamma_{\rm{single}}$ and $\Gamma_{\rm{pair}}$. 
A puzzling feature of the cuprate specific heat jump is its non-mean-field-like evolution with magnetic field\cite{JUNOD93,INDERHEES}. Rather than shifting to lower temperatures, it broadens and reduces in amplitude with little or no change in onset temperature. The fits explain this in terms of an increase in $\Gamma_{\rm{pair}}$ with field, without requiring a reduction in gap magnitude. Note that taking $\Delta(H)=\Delta_0\sqrt{1-(H/H_{\rm{c2}})^2}$ from Ginzburg-Landau theory\cite{DOUGLASS}, the estimated reduction in the gap at 13 T near $T_{\rm c}$ is only 7 to 2 percent for upper critical fields in the range 50 to 100 T. Other properties will now be calculated using the parameters in figure~\ref{FITFIG}.

\subsection{Superfluid Density}
The two scattering rates, $\Gamma_{\rm{pair}}$ and $\Gamma_{\rm{single}}$ are inserted into the anomalous Green's function $F$ as follows
\begin{equation}
F(\xi,\omega)=\frac{\Delta}{(\omega+\xi+i\Gamma_{\rm{pair}})\left(\omega-\xi+i\Gamma_{\rm{single}}-\frac{\Delta^2}{\omega+\xi+i\Gamma_{\rm{pair}}}\right)}
\label{eq:F}
\end{equation}
The superfluid density $\rho_{\rm s}$ is proportional to the inverse square of the penetration depth ($\lambda$) calculated from\cite{CARBOTTE}
\begin{equation}
\fl\frac{1}{\lambda^2(T)} = \frac{16\pi e^2}{c^2V}\sum_\textbf{k}{v_x^2}\int{d\omega^\prime d\omega^{\prime\prime}\lim_{q \rightarrow 0}\left[\frac{f(\omega^{\prime\prime})-f(\omega^\prime)}{\omega^{\prime\prime}-\omega^\prime}\right]} \times B(\textbf{k+q},\omega^\prime)B(\textbf{k},\omega^{\prime\prime})
\label{eq:lsq1}
\end{equation}
where the anomalous spectral function $B$ is given by the imaginary part of $F$.
For a free-electron-like parabolic band $\xi(\textbf{k})=\hbar^2(k_x^2+k_y^2)/2m-\mu$,  $v_x=\hbar k_x/m = \sqrt{2(\xi+\mu)/m}\cos\theta$ and changing variables to $\xi$ and $\theta$ gives
\begin{equation}
\fl\frac{1}{\lambda^2(T)} \propto \int{(\xi+\mu)\cos^2\theta}\int{\left[\frac{f(\omega^{\prime\prime})-f(\omega^\prime)}{\omega^{\prime\prime}-\omega^\prime}\right]} \times B(\xi,\theta,\omega^\prime)B(\xi,\theta,\omega^{\prime\prime})d\omega^\prime d\omega^{\prime\prime}d\xi d\theta
\label{eq:lsq2}
\end{equation}
\begin{figure}
\centering
\includegraphics[width=13cm]{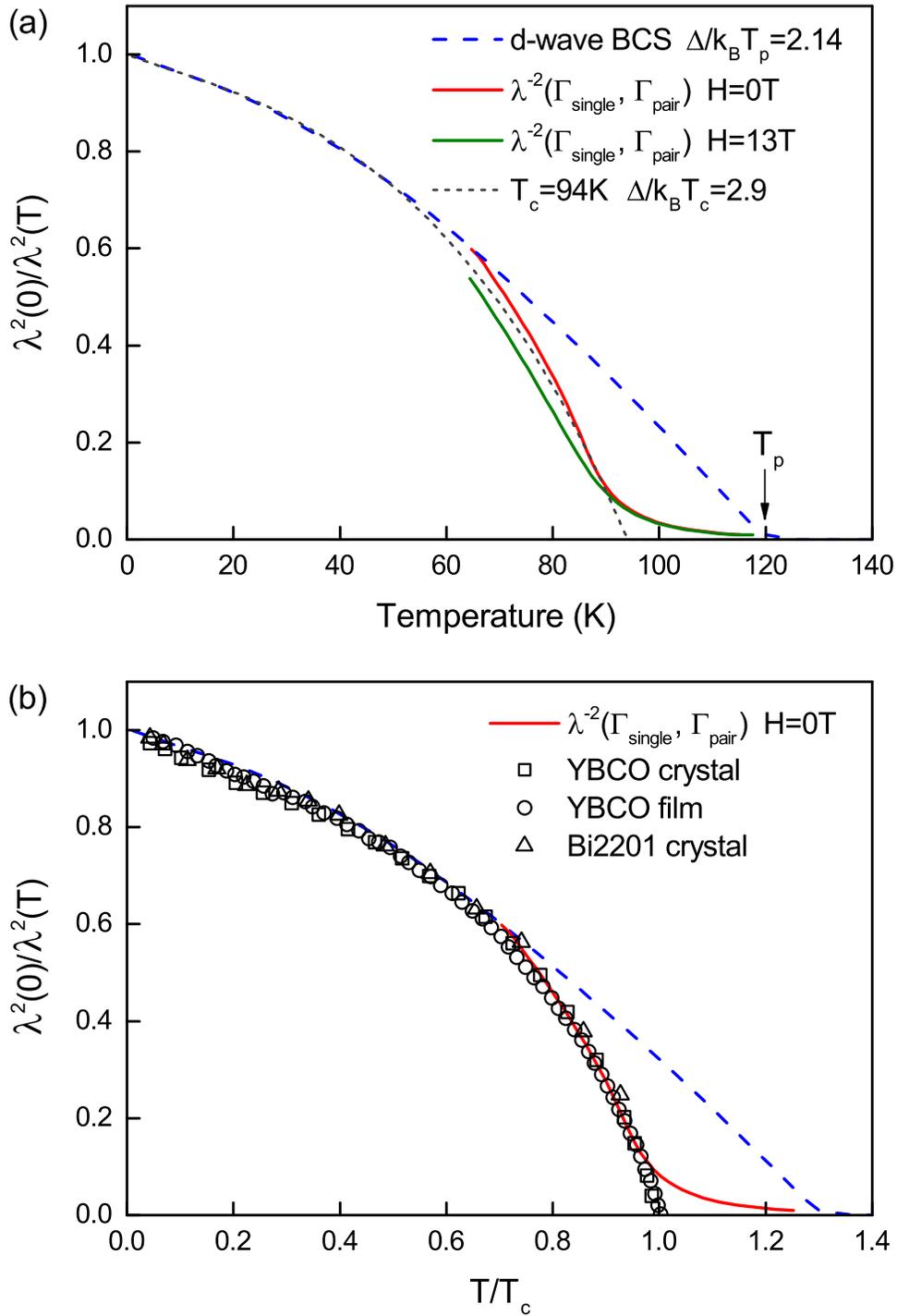}
\caption{
(a) Normalized superfluid density calculated using the parameters from fits to the specific heat in figure~\ref{FITFIG} (red and green lines). (b) Comparison of the calculated zero-field superfluid density with experimental data from \cite{HARDY,BOYCE,KHASANOV2009}.
} 
\label{RHOSFIG}
\end{figure}

The $T$-dependence of $\Gamma_{\rm{pair}}$ causes a clear steepening of $\rho_{\rm s}$ away from the BCS $T$-dependence, with the main onset being pushed down from $T_{\rm p}$ to $T_{\rm c}$, see figure~\ref{RHOSFIG}(a). The same result can be obtained using one scattering rate equal to the average of $\Gamma_{\rm{single}}$ and $\Gamma_{\rm{pair}}$ at each temperature. When plotted in terms of reduced temperature $T/T_{\rm c}$, there is a very good match with experimental data from optimally doped cuprates (figure~\ref{RHOSFIG}(b)). The data, taken by different techniques, includes a YBa$_2$Cu$_3$O$_{7-\delta}$ (YBCO) crystal\cite{HARDY} and film\cite{BOYCE} with $T_{\rm c}$'s near 90 K, as well as a (BiPb)$_2$(SrLa)$_2$CuO$_{6+\delta}$ crystal\cite{KHASANOV2009} with a $T_{\rm c}$ of 35 K. This raises the question as to whether the mooted Berezinskii-Kosterlitz-Thouless universal jump in superfluid density may not simply be attributable to the rapid increase in pair breaking scattering rate near $T_{\rm c}$ arising from fluctuations on a pairing scale that exceeds $T_{\rm c}$\cite{HETEL}. Although the tail above $T_{\rm c}$ is not evident in the selected experimental data, it is observed elsewhere in the literature\cite{JACOBS}. There is a resemblance to an approximate strong-coupling $T$-dependence (dotted line in figure~\ref{RHOSFIG}(a)), calculated from a rescaled BCS gap of magnitude $\Delta_0 = 2.9k_{\rm B}T_{\rm c}$ closing at $T_{\rm c}$ = 94 K, in the absence of strong pair-breaking. However, as will be seen in the following sections, this interpretation of $\rho_{\rm s}(T)$ is inconsistent with other observations.
The suppression in superfluid density with field bears a qualitative similarity to field-dependent measurements on a YBCO thin film\cite{PESETSKI}, but because of that sample's apparent low upper critical field the calculated suppression is much smaller in magnitude.

\subsection{Tunneling}
The current-voltage curve for a superconductor-insulator-superconductor (SIS) tunnel junction is calculated from\cite{MIYAKAWA98}
\begin{equation}
I(V) \sim \int{g(E)g(E-eV)[f(E)-f(E-eV)]dE}
\label{eq:SIS}
\end{equation}
where $g(E)$ is the density of states given by \ref{eq:DOS}. 
\begin{figure}
\centering
\includegraphics[width=\linewidth]{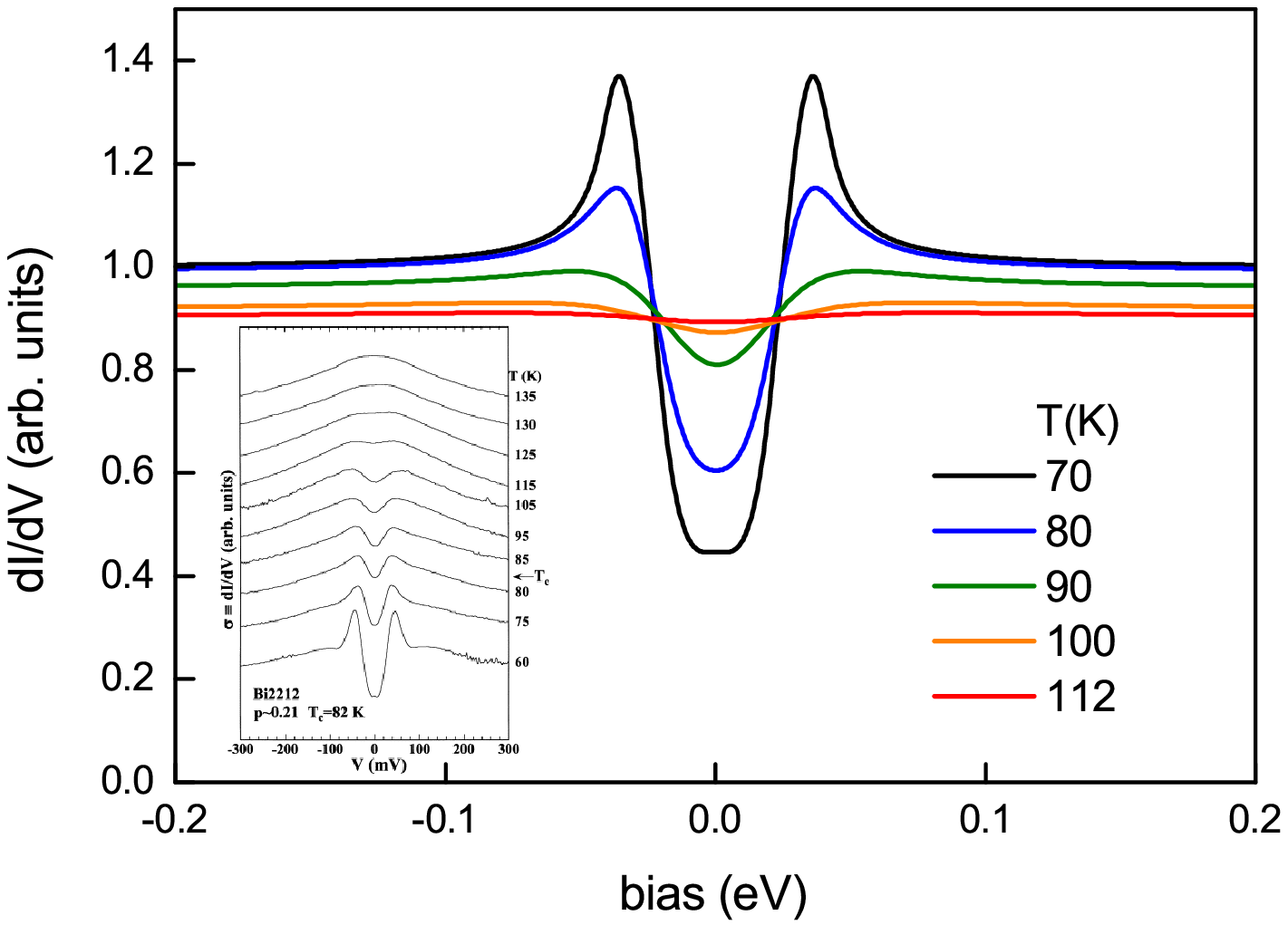}
\caption{
SIS junction tunneling conductance at several temperatures around $T_{\rm c}$ calculated using the parameters from fits to the specific heat in figure~\ref{FITFIG}. Inset: Experimental data reproduced from \cite{DIPASUPIL}.
} 
\label{SISFIG}
\end{figure}
The tunneling conductance $dI/dV$ is plotted in figure~\ref{SISFIG} for several temperatures around $T_{\rm c}$.
The evolution of the spectra with temperature is very consistent with experimental observations\cite{DIPASUPIL,OZYUZER,REN2012,RENPRB2012,BENSEMAN}. These show a filling-in of the gap with temperature and a broadening and suppression of the peaks at $2\Delta$, with little or no shift in their positions. This is contrary to the expected shift toward zero voltage that would occur for a strong coupling gap closing at $T_{\rm c}$ in the absence of pair-breaking scattering.
A depression persists above $T_{\rm c}$ and vanishes as $T_{\rm p}$ is approached, where the superconducting gap closes. Remember that a pseudogap is not included in these calculations. The linearly sloping background seen in the experimental data can be reproduced by adding a linear-in-frequency term, as seen in ARPES\cite{KAMINSKI2005}, to $\Gamma_{\rm{single}}$.

\subsection{Raman Spectroscopy}
Another property that supports the persistence of $\Delta$ above $T_{\rm c}$ is the Raman $B_{\rm{1g}}$ response given by\cite{VALENZUELA}
\begin{eqnarray}
\fl\chi^{\prime\prime}(\omega)=\sum_k{(\gamma_\textbf{k}^{B_{\rm{1g}}})^2\int{\frac{d\omega^\prime}{4\pi}[f(\omega^\prime)-f(\omega^\prime+\omega)]}}
\nonumber\\
\times[A(\textbf{k},\omega^\prime+\omega)A(\textbf{k},\omega^\prime)-B(\textbf{k},\omega^\prime+\omega)B(\textbf{k},\omega^\prime)]
\label{eq:Raman1}
\end{eqnarray}
The Raman $B_{\rm{1g}}$ vertex $\gamma_\textbf{k}^{B_{\rm{1g}}} \propto \cos k_x-\cos k_y \sim \cos 2\theta$ probes the antinodal regions of the Fermi surface where $\Delta(\textbf{k})$ is largest. Changing variables from $\textbf{k}$ to $\xi$ and $\theta$ gives
\begin{eqnarray}
\fl\chi^{\prime\prime}(\omega)\propto\int{d\xi d\theta \cos^2(2\theta)\int{d\omega^\prime}[f(\omega^\prime)-f(\omega^\prime+\omega)]}
\nonumber\\
\times[A(\xi,\theta,\omega^\prime+\omega)A(\xi,\theta,\omega^\prime)-B(\xi,\theta,\omega^\prime+\omega)B(\xi,\theta,\omega^\prime)]
\label{eq:Raman2}
\end{eqnarray}
\begin{figure}
\centering
\includegraphics[width=13cm]{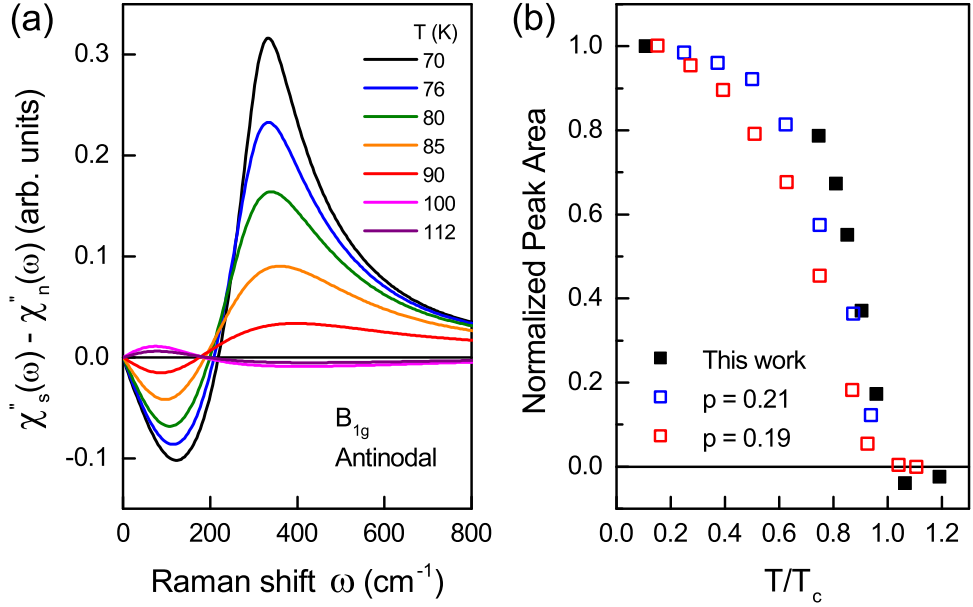}
\caption{
(a) Difference between the superconducting and normal-state (i.e. just above $T_{\rm p}$) antinodal ($B_{\rm{1g}}$) Raman response functions at temperatures around $T_{\rm c}$, calculated using the parameters from fits to the specific heat in figure~\ref{FITFIG}.
(b) Normalized area under the curves in (a) compared with experimental values from \cite{BLANC} for dopings $p$=0.19 and 0.21.
} 
\label{B1GFIG}
\end{figure}
The superconducting Raman $B_{\rm{1g}}$ response function with the normal-state response at 122 K subtracted is shown in figure~\ref{B1GFIG}(a) for several temperatures around $T_{\rm c}$.
The resemblance to experimental data, reported in \cite{GUYARD,GUYARDPRL08,BLANC}, is striking. Like the tunneling results above, the peak at $2\Delta$ broadens and reduces in amplitude and barely shifts with temperature indicating that the gap magnitude is still large at $T_{\rm c}$\cite{GUYARD}. Figure~\ref{B1GFIG}(b) shows the normalized area under the curves in (a) versus reduced temperature ($T/T_c$=94K), together with data from \cite{BLANC}. The calculations show that data plotted in this way  gives little indication of a gap above $T_c$.

\subsection{Optical Conductivity}
The final property considered in this work is the $ab$-plane optical conductivity calculated from\cite{YANASE} 
\begin{eqnarray}
\fl\sigma(\omega)=\frac{e^2}{\omega}\sum_\textbf{k}{v_{\rm{ab}}^2(\textbf{k})\int{\frac{d\omega^\prime}{\pi}[f(\omega^\prime)-f(\omega^\prime+\omega)]}}
\nonumber\\
\times[A(\textbf{k},\omega^\prime)A(\textbf{k},\omega^\prime+\omega)+B(\textbf{k},\omega^\prime)B(\textbf{k},\omega^\prime+\omega)]
\label{eq:Opcon1}
\end{eqnarray}
where $v_{\rm{ab}}(\textbf{k})=\sqrt{v_x^2+v_y^2}$. Again a change of variables is made from momentum to energy and Fermi surface angle as follows
\begin{eqnarray}
\fl\sigma(\omega)\propto\frac{1}{\omega}\int{d\xi d\theta(\xi+\mu)\int{d\omega^\prime[f(\omega^\prime)-f(\omega^\prime+\omega)]}}
\nonumber\\
\times[A(\xi,\theta,\omega^\prime)A(\xi,\theta,\omega^\prime+\omega)+B(\xi,\theta,\omega^\prime)B(\xi,\theta,\omega^\prime+\omega)]
\label{eq:Opcon2}
\end{eqnarray}
\begin{figure}
\centering
\includegraphics[width=13cm]{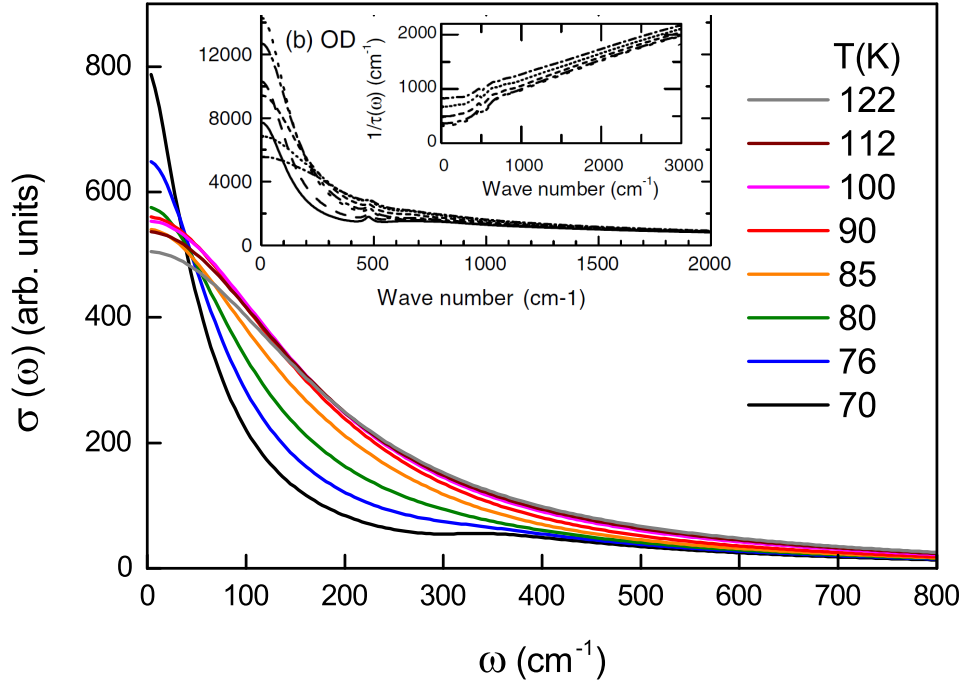}
\caption{
$ab$-plane optical conductivity at several temperatures around $T_{\rm c}$ calculated using the parameters from fits to the specific heat in figure~\ref{FITFIG}.
Inset: Experimental data reproduced from \cite{SANTANDER}
} 
\label{OPCONFIG}
\end{figure}
Spectra at several temperatures around $T_{\rm c}$ are plotted in figure~\ref{OPCONFIG}. A suppression is visible below $2\Delta$ at low temperature that fills in as temperature is increased. A gap closing at $T_{\rm c}$ would result in the onset of this suppression shifting to lower frequency. The calculations bear a strong qualitative resemblance to the overdoped data reported by Santander-Syro \textit{et al}.\cite{SANTANDER}

\section{Discussion}
As summarized in table~\ref{FLUCTABLE} only the superfluid density, and more approximately the zero-field specific heat, can be interpreted by a strong-coupling gap closing at $T_{\rm c}$ in the absence of scattering. The non-mean-field $T$-dependence of all properties examined in this work is instead well described in terms of a superconducting gap that persists above $T_{\rm c}$, in the presence of a steep increase in scattering. This result is insensitive to the addition of linear-in-frequency terms or a $\cos2\theta$ momentum dependence to $\Gamma_{\rm{pair}}$ and $\Gamma_{\rm{single}}$. The scattering is further enhanced by magnetic field. What is the origin of the scattering and can it be suppressed to bring $T_{\rm c}$ up to $T_{\rm p}$? A rapid collapse in quasiparticle scattering below $T_{\rm c}$, also found in microwave surface impedance measurements\cite{BONN}, is expected when inelastic scattering arises from interactions that become gapped or suppressed below $T_{\rm c}$\cite{HOSSEINI}. The spin fluctuation spectrum is a plausible candidate and has been investigated extensively\cite{QUINLAN,DUFFY}, although those calculations assumed that the superconducting gap closes at $T_{\rm c}$.
\begin{table}
\caption{Interpretations of the experimental data.}
\centering
\begin{tabular}{lcc}
\br
Technique & $\Delta$ closing at $T_{\rm c}$ & $\Delta$ above $T_{\rm c}$ \\
& (strong coupling) & plus scattering\\
\mr
Specific heat &\checkmark &\checkmark\\
Penetration depth &\checkmark &\checkmark\\
ARPES & &\checkmark\\
Tunneling & &\checkmark\\
Raman spectroscopy &  &\checkmark\\
Optical conductivity & &\checkmark\\
\br
\end{tabular}

\label{FLUCTABLE}
\end{table}

The work presented here illustrates that the merging of the $T^*$ line on the lightly overdoped side of the $T_{\rm c}$ dome is not a product of the pseudogap per se, but rather the persistence of the superconducting gap into the fluctuation region between $T_{\rm c}$ and $T_{\rm p}$. As doping increases, this region becomes narrower and experimental properties become more mean-field-like. Switching direction, as doping decreases the pseudogap opens, grows, and eventually exceeds the magnitude of the superconducting gap at the antinodes. When this occurs, the gap associated with $T^*$ changes to the pseudogap. In other words, $T^*$ is given by the larger of $T_{\rm p}$ and $E_g/2k_{\rm B}$ (see figure~\ref{TSTARFIG}(c)). Such an interpretation makes immediate sense of the phase diagram presented by Chatterjee \textit{et al}.\cite{CHATTERJEE}

\ack
Supported by the Marsden Fund Council from Government funding, administered by the Royal Society of New Zealand. The author acknowledges helpful discussion with J.L. Tallon.

\end{document}